\documentstyle[12pt]{article}
\newcommand{\nl}{\nonumber \\}
\newcommand{\be}{\begin{equation}}
\newcommand{\ee}{\end{equation}}
\newcommand{\ba}{\begin{eqnarray}}
\newcommand{\ea}{\end{eqnarray}}
\newcommand{\ci}[1]{\cite{#1}}
\newcommand{\bi}[1]{\bibitem{#1}}
\newcommand{\la}[1]{\label{#1}}

\newcommand{\gmnu}{g^{\mu\nu}}
\newcommand{\gmnd}{g_{\mu\nu}}
\newcommand{\wt}{\widetilde}

\date{}
\begin{document}
\begin{center}
{\Large
Chiral Bosonization of $U_A(1)$-Currents and the Energy-Momentum Tensor
in Quantum Chromodynamics
\footnote{Recently the new interest arised \cite{arr} 
to the chiral coefficients 
$L_{11}, L_{12},  L_{13}$ in the low-energy ChPT lagrangian describing
the pseudoscalar meson coupling to gravity and/or light singlet
dilatons. In 90ties we estimated these coefficients within the 
Chiral and Conformal Bosonization Model \cite{Andr2,Andr3} but published the
results in the journal \cite{zap} not easily accessible for the physics community. 
 We put now our old paper to the E-archives. }'\footnote{
This is a translation of the article published
in "Zapiski nauch. sem. POMI" (Proc. Steklov Math. Inst.,St.Petersburg branch)
 v.224/13 (1995) 68; Engl.transl.\cite{zap};  
 its short version appeared in \cite{aayu}.}}

\vspace{0.4cm}

{\bf\large Andrianov A.A., Andrianov V.A. ,Yudichev V.L.\\}
{\sl V.A.~Fock Department of Theoretical Physics,
St.-Petersburg State University,\\
198504 St.-Petersburg, Russia}
\vspace{0.4cm}
\end{center}
\begin{flushright}
\it Dedicated to Memory of Victor N. Popov
\end{flushright}

\section{Introduction.}
\hspace*{5mm}In the various fields of physics, in particular,
in description of quantum phenomena at
low energies and temperatures, there appears the problem of
extraction of essential (collective) variables and construction of a relevant
effective lagrangian \ci{Popov}.
Our interest to these problems had arisen from the numerous discussions
with Victor N. Popov, whose experience and contribution to this domain
helped us to develop the method of the low-energy bosonization \cite{Andr1}
and apply it to the
theory of strong interactions --- Quantum Chromodynamics .

 Chiral lagrangians for the light
$SU(3)$ pseudoscalar mesons parameterize their
strong and electroweak interactions by means of more than ten
constants \cite{GasLeu},
which contain information about the dynamics of
 the quark and gluon interaction
--- Quantum Chromodynamics (QCD).
Most of the constants can be evaluated
from the analysis of experimental data,
thereby there appears a possibility
to examine QCD at low energies.
For the  calculation of the structure constants
of the chiral lagrangian
the low-energy  QCD bosonization
method has been developed \cite{Andr1}.
This method gives good numerical results for the $SU(3)$
quark current matrix elements \cite{Andr3}.

Recently, the necessity to extend
the $SU(3)_F$-chiral lagrangian appears 
to describe $U_A(1)$ current matrix elements
\ci{DonWyl}, in particular, for
the pseudoscalar gluon density, and also
the  matrix elements of the energy-momentum tensor \cite{DonLeu} ,
amonge them,for  the scalar gluon density.
These matrix elements characterize some decays of heavy quarkonium  ($\bar
cc, \bar bb $ and etc. ) \ci{DonWyl} and also Higgs bosons,
which occur via so-called vacuum channel.
Thus, it is interesting to apply the low-energy  QCD-bosonization method
for the construction of the generalized chiral lagrangian in the presence of
 external $U_A(1)$-fields and in the background metric $g_{\mu\nu}(x)$.

The main goal of this work is to  elaborate the
parameterization of the extended chiral lagrangian
by means of the low-energy  QCD-bosonization method, in
the model \ci{Andr1}, and derivation of the
relations between the chiral coupling constants
in the limit of large-$N_c$, number of colors ( generalized Zweig rules ).

In the section 2 we briefly explain the scheme of the low-energy
QCD-bosonization, display the structure of the $SU(3)_F$-chiral
lagrangian in the presence of the external vector, axial-vector, scalar
and pseudoscalar fields, give  estimates of its structure
constants within the model \ci{Andr1}.  Further on, following the
ideology of  the Chiral Perturbation Theory \ci{GasLeu}, we transform the
bosonized chiral lagrangian into a phenomenological one and find 
relations (of Zweig rule type) between the corresponding structure
constants.  In the section 3 the $U(3)$-chiral
bosonization in the presence of external
$U_A(1)$-fields is considered in the heavy
$\eta'$-meson mass limit.  It makes possible to obtain the
$SU(3)$ chiral lagrangian in the presence of external $U(3)$ fields
and estimate new structure constants $L_{14}...L_{19}$.  The
corresponding vertices describe the  heavy hadron decays
which occur via the vacuum pseudoscalar channel.  
The generalized Zweig rules are found that allow to express the new structure
coupling constants through the known ones.  In the section 4
the conformal anomaly of the quark determinant \ci{Andr2} is
used for the construction of the chiral lagrangian in external
conformal-flat metric of the space-time.  It turns out to be
sufficient to determine three structure constants of the
generalized chiral lagrangian in the vertices characterizing
the meson energy-momentum tensor matrix elements.
In the conclusion we discuss the correspondence of the predictions
obtained  by  the low-energy QCD-bosonization method and  to other
theoretical
and phenomenological estimates.

\section{
SU(3)-Chiral Lagrangian in the Low-Energy Bosonization Method.}

\hspace*{5mm}Let us expose the scheme of
the low-energy  QCD bosonization for
the $ SU(3) $-chiral lagrangian. The QCD-bosonization is carried out
by transition from the color quark and gluon fields to the collective
boson variables describing the pseudogoldstone  excitations which appear
after the dynamical chiral symmetry breaking.
 For the  generating
functional of the singlet quark current correlators
$ Z(V,A,S,P) $ the bosonization
is given by the following identity:

\be
\int\!{\cal D}G\,{\cal D}\psi{\cal D}\bar \psi
\exp (- S_{QCD}(\bar \psi,\,\psi,\,G;\,\,V, A, S, P))
= \int\! {\cal D}U
\exp(- S_{eff}(U;\,\, V, A, S, P)) \cdot {\cal R}, \la{genfunc}
\ee
where
$ V,A,S,P $
are, respectively, the external vector, axial-vector, scalar and
pseudoscalar sources, and
$ U(x) $ is parametrized by pseudoscalar fields:

\be
U=\exp (i\Pi (x)/F_{\pi})\quad F_{\pi}\approx 93 MeV.
\ee

In the right hand side of (\ref{genfunc}) the integration is made over the
finite number of light boson states, and the heavier states'
effects are included in the discrepancy functional
$R $.
The optimal bosonization at low energies  guarantees
$ R\approx 1 $.
It corresponds to taking into account the total highest excitations' effects
in the structure constants of chiral lagrangian.
In the low-energy  bosonization model \ci{Andr1} the dynamical chiral
symmetry breaking in the low-energy  region {\bf L} is approximated by
two parameters:
$ \Lambda $, the top spectrum boundary, and
$ M $, the spectrum asymmetry of the dynamical quarks.
In this region
{\bf L}$=(\Lambda, M) $ the choice of boson variables is based on the
chiral noninvariance of the generating functional  under the local chiral
transformations of external fields (chiral anomaly).
The derivation of the effective chiral action is produced by integration
of the generating functional over the group of local chiral rotations:

\be
Z_{inv}^{-1}=\int {\cal
D}U Z_\psi^{-1}(V^U,A^U,S^U,P^U) ,   \la{quarkdet}
\ee
where $ Z_\psi $ is a quark part of the generating functional
before integrating over gluons. The integration in
(\ref{quarkdet}) is carried out over the local $ SU(3) $ group
with invariant measure.  Thus, we  obtain a chiral invariant
part of the generating functional $ Z $.  Due to this, one
succeeds  to calculate the chiral noninvariant part which is
the  effective action for the pseudoscalar fields, $
S_{eff}(U;V,A,S,P) $:

\be
Z=\int {\cal D}U\frac{Z}{Z_\psi(V^U,A^U,S^U,P^U)}<\!<Z_{inv}>\!>_G
\equiv \int {\cal D}U \exp (i S_{eff}(U;V,A,S,P))<\!<Z_{inv}>\!>_G,
\la{bos sc}
\ee
where $<\!<...>\!>_G$ means  averaging over gluons.

In the model \ci{Andr1}
$ S_{eff}$ does not contain gluon fields, what is reflected in (\ref{bos sc}).
The role of gluons is reduced only to the formation of the
dimensional parameters of the theory,
{\bf L}$=(\Lambda, M) $, on the base of the equation of stability of
the low-energy  region
, formed by gluon condensate \ci{Andr3}.

In the low-energy region {\bf L} the vertices of chiral lagrangian  are
classified by momentum and pseudoscalar meson mass degrees
in accordance with the Chiral Perturbation Theory rules
\ci{GasLeu}:

\be
S_{eff} = \int d^4x ({\cal L}_2  + {\cal L}_4) \,+\, S_{WZW}.
\ee
The Weinberg lagrangian of dimension 2, ${\cal L}_2$, looks as:
\be
{\cal L}_2 = \frac{F^2_0}4[<(D_{\mu}U)^ + D^{\mu}U>
+ <{\chi}^+U + U^+\chi >],
\ee
where the symbol
$ <...> $ means averaging over flavor indices, the external
sources are included in the covariant derivative
$D_{\mu}=\partial_{\mu}+[V_{\mu},*]+\{A_{\mu},*\}$
and in the complex density
$\chi =2B_0(S+iP)$ with  the parameter
$ B_0 $ connected with the condensate
$<\bar\psi\psi>=-F_0^2B_0$.
In \ci{Andr1} these parameters are determined by model
characteristics in the low-energy region L:

\be
F^2_0=\frac{N_c}{4\pi^2}(\Lambda^2-M^2),\qquad F_0^2B_0=\frac{N_c}{2\pi^2}
(\Lambda^2M-\frac13 M^3).
\ee
The Wess-Zumino-Witten action,
$ S_{WZW} $, contains abnormal parity vertices, its form
presented in \ci{Andr1}, and in context of this paper
it  turns out to be important for the
$ U_A(1) $-bosonization (see the nest section).

The lagrangian of dimension 4, ${\cal L}_4$, responds for the fine
structure of interaction of the pseudoscalar mesons and includes
nine structure constants $ I_k $:

\ba
{\cal L}_4^{eff} &=&
I_1<D_{\mu}U(D_{\nu}U)^\dagger D^{\mu}U(D^{\nu}U)^\dagger > +
 I_2<D_{\mu}U(D_{\mu}U)^\dagger D^{\nu}U(D^{\nu}U)^\dagger > \nl
&&+ I_3 <(D_{\mu}^2U)^\dagger D_{\nu}^2U>+
I_4<(D_{\mu}\chi)^\dagger D^{\mu}U + D_{\mu}\chi (D^{\mu}U)^\dagger > \nl
&&+ I_5 <D_{\mu}U(D^{\mu}U)^\dagger (\chi
U^{\dagger}  + U\chi{^\dagger}) >
+ I_6 <U{\chi}^\dagger U{\chi}^\dagger  + \chi U^\dagger \chi U^\dagger > \nl
&&+ I_7<{\chi}^\dagger U - U^\dagger \chi >^2\nl
&&+ I_8<F_{\mu \nu}^R D^{\mu}U(D^{\nu}U)^\dagger  + F_{\mu
\nu}^L(D^{\mu}U)^\dagger D^{\nu}U> + I_9<U^\dagger F_{\mu
\nu}^R U F^{L \mu \nu}>,      \label{efflag}
\ea
where
$F^L_{\mu\nu}=\partial_{\mu}L_{\nu}-\partial_{\nu}L_{\mu}+[L_{\mu},L_{\nu}];\qquad L_{\mu}=V_{\mu}-A_{\mu};\quad R_{\mu}=V_{\mu}+A_{\mu}$.
After the bosonization one obtains the following estimates for the
constants
$ I_k $:

\ba I_1=\frac{N_c}{192\pi^2},\qquad
I_2=-\frac{N_c}{96\pi^2},\qquad I_3=\frac{N_c}{96\pi^2},\qquad
I_4=\frac{N_cM}{16\pi^2B_0},\qquad I_5=0, \nl
I_6=-\frac{\Lambda^2-M^2}{64\pi^2B_0},\qquad I_7=0,\qquad
I_8=\frac{N_c}{48\pi^2},\qquad I_9=\frac{N_c}{96\pi^2}.
\la{estim}
\ea
From (\ref{estim}) one can see that the
$ SU(3) $-low-energy bosonization  reproduces  the
 coefficients $ I_k $ in the main order of
$N_c\quad (I_k=O(N_c),\quad k\not=7)$; the coefficient
$ I_7 $  is saturated by the gluonic vacuum pseudoscalar configurations
and is not determined correctly in the
$ SU(3)$-bosonization (see the next section).

The phenomenological chiral lagrangian can be found from the lagrangian
(\ref{efflag}) by rules of the Chiral Perturbation Theory, when the
equation of motion for the lagrangian of dimension 2 is use:

\be
 (D_{\mu}^2U)^{\dagger }U-U^{\dagger }D_{\mu}^2U-{\chi}^{\dagger }U
 + U^{\dagger }\chi
=\frac{1}{3}< U^{\dagger }\chi -{\chi}^{\dagger }U>,
\la{moteq}
\ee
In the $ SU(3) $ case, the following identities  turn out to be useful:
\be
<  A_{\mu}A_{\nu}A_{\mu}A_{\nu} > =
- 2 <  A_{\mu}^2 A_{\nu}^2 > + \frac{1}{2}
<  A_{\mu}^2 >^2 + <  A_{\mu}A_{\nu} >^2
\ee
The additional Gasser-Leutwyler lagrangian of dimension 4 \ci{GasLeu}
contains ten structure constants
$ L_i\quad (i=1,...,10) $:

\ba
{\cal L}_4^{GL}&=&
L_1< (D_{\mu}U)^{\dagger }D^{\mu}U>^2 +
L_2< (D_{\mu}U)^{\dagger }D_{\nu}U>\!
< (D^{\mu}U)^{\dagger }D^{\nu}U> \nl
&&+ L_3< (D^{\mu}U)^{\dagger }D_{\mu}U(D^{\nu}U)^{\dagger }
D_{\nu}U> +
L_4< (D^{\mu}U)^{\dagger }D_{\mu}U>< {\chi}^{\dagger }U
+ U^{\dagger }\chi> \nl
&&+ L_5< (D^{\mu}U)^{\dagger }D_{\mu}U({\chi}^{\dagger }U
+ U^{\dagger }\chi)>
 + L_6< {\chi}^{\dagger }U + U^{\dagger }\chi >^2 \nl
&& + L_7< {\chi}^{\dagger }U - U^{\dagger }\chi >^2 +
L_8< {\chi}^{\dagger }U{\chi}^{\dagger }U +
U^{\dagger }\chi U^{\dagger }\chi >\nl
&&+ L_9< F_{\mu \nu}^R D^{\mu}U(D^{\nu}U)^{\dagger } + F_{\mu
\nu}^L(D^{\mu}U)^{\dagger }D^{\nu}U> - L_{10}< U^{\dagger }F_{\mu
\nu}^R U F^{L \mu \nu}>,  \la{lag3}
\ea
which  turn out to be connected with the coefficients
$ I_k $ in the chiral bosonization models:

\ba
2L_1&= & L_2= I_1,\quad L_3 =  I_2 + I_3 - 2I_1,
\quad L_4 = L_6 = 0,\quad L_5 = I_4 + I_5,\nl
L_7 &=& I_7 -\frac{1}{6} I_4 + \frac{1}{12} I_3,
\quad L_8 =  -\frac{1}{4} I_3 + \frac{1}{2} I_4 + I_6,\quad
L_{9} = I_8,\quad L_{10} =  I_9 .
\la{Zw0}
\ea
The zero-valued
$ L_4, L_6 $ are in compliance with the large-$ N_c $ counting rules,
since they give contribution of order $O(1)\quad (N_c\rightarrow \infty)$.
This choice eliminates the uncertainty characteristic for the
$ SU(3) $-lagrangians of dimensions 2 and 4, which appears due to
the symmetry of observables under the one-parameter transformation:

\be
\chi^{(\lambda)}=\chi+\frac{\lambda}{B_0^2}(\det\chi^+)\chi(\chi\chi^+)^{-1}
\ee
(Kaplan-Manohar symmetry).
Here
$ \lambda $ is an arbitrary real number.
Such a transformation changes values of the structure constants
$L_6,L_7,L_8$ as follows:

\be
L_6^{\lambda}=L_6-\tilde\lambda,\qquad
L_7^{\lambda}=L_7-\tilde\lambda,\qquad L_8^{\lambda}=L_8+2\tilde\lambda,
\ee
where  $\tilde\lambda=F_{\pi}^2\lambda/16B_0^2$.
Setting
$ L_6=0 $ in  accordance with the Zweig rules, we evidently fix the constants
$ L_7,L_8 $.

So far as the structure of the chiral lagrangian, presented in
(\ref{efflag}), is general for the QCD-bosonization models in
the limit of large number of colors, the correlations
(\ref{Zw0}) between the model coupling constants $ I_k $ and
the phenomenological ones $ L_k $ characterize an uncertainty
in fixing of the constants $ I_k $ from the experimental data.
There exists a two-parameter family of chiral bosonization
models, giving the same values of observables.

\section{
$ U_A(1) $ Generalization of the Chiral Lagrangian.
}

\hspace*{5mm}The chiral lagrangian in the presence of external fields,
singlet in flavors, allows to conduct the bosonization of the
singlet quark currents and pseudoscalar gluon density.  To
construct it correctly, it is necessary to extend the $
SU(3)$-lagrangian up to the $ U(3)$-type one and take into
account correctly the influence of the vacuum QCD effects.
The external $ U_A(1)$-fields include the axial-vector singlet
field $ A_{\mu} $ and the source $ \theta $ coupling with the
pseudoscalar gluon density $G\widetilde G$:

\be
{\cal L}_{\theta} = - \theta (x) G \widetilde
G,\quad\mbox{где}\quad G \widetilde G \equiv
 \frac{\alpha_s}{8\pi} G^a_{\mu \nu}\widetilde
G^{\mu \nu}_a . \la{an1}
\ee
Let us accomplish the generalized scheme of the low-energy
bosonization for the $ U(3) $ case and consider the
chiral fields $U(x)\quad\longrightarrow\quad\wt U(x)$, $\wt U=U
exp(i\eta_0/3)$ as collective variables.  In general case, it gives
the additional term in the lagrangian:

\be
{\cal L}_- (\eta_0) = \Biggl(\eta_0 + i \xi <  \chi^{\dagger }
\widetilde U - \widetilde U^{\dagger } \chi>\Biggr) G \widetilde G,
\la{an2}
\ee
which is conditioned by the chiral anomaly of the quark determinant.
In accordance with the counting rules when $N_c$ is large,
$\xi=O(1) $. When comparing the vertices (\ref{an1}) and (\ref{an2})
we see that the external source $\theta$ can be combined with
the scalar density $\chi\longrightarrow
\tilde\chi=\chi\exp(i\theta(x)/3)$, if change the
variable $\eta_0\longrightarrow\eta_0-\theta$.
After this, in the $SU(3)$ lagrangian there appear the new vertices with six
phenomenological constants \ci{DonWyl}:

 \ba {\cal L}_4^{(D\theta )}&= &  iL_{14}D_{\mu}D^{\mu}\theta
 < {\widetilde \chi}^{\dagger } U - U^{\dagger } \widetilde\chi> +
 iL_{15}D_{\mu}\theta < (D^{\mu} \widetilde\chi)^{\dagger } U
 - U^{\dagger }(D^{\mu} \widetilde\chi )> \nl
 &+& L_{16} D_{\mu}\theta D^{\mu}\theta
  < D_{\nu}U(D^{\nu}U)^{\dagger }> +
 L_{17} D_{\mu}\theta D_{\nu}\theta
  < D^{\mu}U(D^{\nu}U)^{\dagger }> \nl
 &+& L_{18}D_{\mu}\theta D^{\mu}\theta
 <\widetilde\chi U^{\dagger } + U \widetilde\chi^{\dagger }>+
 iL_{19}D_{\mu}\theta < U(D^{\mu}U)^{\dagger } D_{\nu}U
 (D^{\nu}U)^{\dagger }>,
\la{DW}
\ea
which in the chiral bosonization models turn out to be connected with
the structure constants of the
$ SU(3) $-lagrangian (\ref{efflag}) and (\ref{lag3}):

\ba
L_{15}&= & - 6 L_{18}  = -\frac{2}{3}(I_4 +
I_5)= -\frac{2}{3} L_5   ;\nl L_{16}&= &  \frac{1}{2} L_{17}  =
 -\frac{1}{6} L_{19} = \frac{2}{9}(I_1 + I_2 + I_3)=  \frac{2}{9}(3L_2
 + L_3) \la{Zw1}
\ea
When forming the constant
$ L_{14} $, vacuum effects play a significant role.
They are connected with that at large distance in QCD there
appear infrared effects leading to non-zero value of the pseudoscalar
gluon density correlator (topolgical susceptibility).

\be
     M_0^4 = -\int
     d^4\!x < 0|T(G\widetilde G(x) G\widetilde G(0))|0>_0. \la{sus}
\ee
This phenomenon guarantees  solution of the so-called
$ U(1) $ problem and accounts for relatively large mass of the
$ \eta'$ meson.
The averaging over gluons with  taking into account (\ref{sus})
and with the multipole expantion approach used
allows to calculate an additional contribution to the
$ U(3) $ meson lagrangian:

\be
\widetilde{\cal L} (\eta_0) =
- \frac{M_0^4}{2} \Biggl(\eta_0 + i \xi <  \chi^{\dagger } \widetilde U
- \widetilde U^{\dagger } \chi>\Biggr)^2+O(\frac{1}{N_c}).
\ee

In this paper we are interested in the bosonization of the QCD-currents
in the light pseudoscalar mesons sector.
Thus, we consider the $\eta'$-meson mass as a large parameter in comparing
to energies characteristic for the other meson's interaction.
It gives us the reason to exclude the $\eta_0$-field by means of
the large mass reduction method, i.e. the $1/M_0$ expansion and the Gaussian
approach.
As a result, we express the remaining constants $L_7, L_{14}$ via
parameters characterizing  interaction in the vacuum
channel.

\be I_7 = -
\frac{1}{2M_0^4}\Biggl(\frac{F_0^2}{12} - \xi M_0^4\Biggr)^2;\quad L_7
= - \frac{F_0^4}{288 M_0^4} -\frac{1}{6} I_4 + \frac{1}{12} I_3 +
\frac{\xi F_0^2}{12} - \frac{\xi^2 M_0^4}{2} .  \la{Zw7} \ee \be L_{14}
= \frac{F_0^4}{72 M_0^4} -\frac{1}{3} I_4 - \frac{2}{3} I_5 - \frac{\xi
F_0^2}{6}  = - 4L_7 + O(N_c), \la{Zw2}
\ee
The terms in these formulas are ordered according to their contributions
at large-$N_c$, namely the first terms in $L_7,L_{14}$ are of order
of $O(N_c^2)$, $N_c=3$, and the next three ones are estimated as
$O(N_c)$.
In this case the Zweig rule works only numerically for $N_c$, and in
the large-$N_c$ limit it is wrong, since the reduction of $\eta'$ by
 $1/M_0$ expansion is invalid.
The saturation of the constants $L_7, L_{14}$ by terms of order $O(N_c)$
is in a good agreement with experimental estimations \ci{DonWyl}:

\be
L_7=(-0.4\pm0.2)\cdot 10^{-3},\qquad L_{14}=(2.3\pm1.1)\cdot 10^{-3}.
\ee

The constraints (\ref{Zw0}) and (\ref{Zw1})
presented above are characteristic
for the most of the chiral bosonization models.
In the chiral bosonization model, presented in the section 2, the parameter
$\xi=0$ that evidently allows to connect  the remaining constants with
the phenomenological ones, i.e. fix them from experiment.
Thus, with  the known values of $\xi$ and $N_c$,  the
uncertainty in the model lagrangian vertices disappears.
Thereby, so-called tachion vertices, which are proportional to
$I_3$ and $I_4$, influence the $U_A(1)$-physics at low energies
and can be estimated from  experimental data.

\section{
Effective Chiral Lagrangian in the Conformal Metric
and the Energy-Momentum Tensor.
}

\hspace*{5mm}The matrix elements of the energy-momentum tensor
 in the chiral theory are shown
to be important for the description of the low-energy processes in which
the scalar  and tensor mesons are involved.
Their calculation can be performed within the Chiral Perturbation
Theory, when a nontrivial background metric in the chiral lagrangian
is introduced.
The metric tensor $g_{\mu\nu}$ is an external source after the variation
of which the energy-momentum tensor can be found:

\be
\frac12\theta_{\mu\nu}(x)=\frac{\partial}{\partial \gmnu(x)}
\sqrt{-g}{\cal L}(\psi,A^{a\mu},\gmnd)\Bigr|_{\gmnd=\eta_{\mu\nu}},
\ee
where $\eta_{\mu\nu}=\mbox{diag}(1,-1,-1,-1)$.

The effective lagrangian is  built of the vertices invariant under
the global coordinate transformations (diffeomorphisms).
The general form of the chiral lagrangian can be written as
follows:

\be
{\cal L}={\cal L}^{(2)}+{\cal L}^{(4)}, \la{genlag}
\ee
where ${\cal L}^{(2)}$, is still the Weinberg lagrangian generalized
to an arbitrary metric.
The $L^{(4)}$ includes terms of order of $p^4$:

\be
{\cal L}^{(4)}={\cal L}^{(4,g)}+{\cal L}^{(4,R)}     .  \la{lag4}
\ee
The first part of $L^{(4,g)}$ in the formula (\ref{lag4}) resembles in one's
structure the
lagrangian (\ref{lag3}) in an arbitrary metric.
The other part, ${\cal L}^{(4,R)}$, includes the curvature tensor:

\ba
{\cal L}^{(4,R)}&= & L_{11}R<D_{\mu}U(D^{\mu}U)^\dagger > \nl
&+ &  L_{12}R^{\mu \nu}<D_{\mu}U(D_{\nu}U)^\dagger > \\
&+ &  L_{13}R<\chi U^\dagger +U{\chi}^\dagger >, \nonumber
\ea
where
$R_{\mu\nu}=R_{\mu\lambda\nu}^{\lambda}$, $R=R_{\mu\nu}\gmnu$,
$ R^{\lambda}_{\mu\sigma\nu}={\partial}_{\sigma}{\Gamma}^{\lambda}_{\nu
 \mu}-{\partial}_{\nu}{\Gamma}^{\lambda}_{\sigma\mu}+
{\Gamma}^{\lambda}_{\sigma\alpha}{\Gamma}^{\alpha}_{\nu\mu}-
{\Gamma}^{\lambda}_{\nu\alpha}{\Gamma}^{\alpha}_{\sigma\mu} $.
Thus, in an arbitrary metric three new phenomenological constants
appear, which prove to be necessary for the energy-momentum
tensor parametrization in the chiral theory

One can convince that for their estimates obtained
 by means of the low-energy
bosonization method it is sufficient to consider the class of conformal metrics.

\be
\gmnd=\exp(-2\sigma)\eta_{\mu\nu},\quad
\eta_{\mu\nu}=\mbox{diag}(1,-1,-1,-1).
 \ee
In this case the model lagrangian with conformal metric is obtaind from
integration over the conformal and chiral anomalies, according to the bosonization
scheme presented in the section 2.

In the conformal-flat metric the scalar curvature $R$ and the Ricci tensor $R_{\mu\nu}$
are functionals depending on $\sigma(x)$ and
$\sigma_{\mu}\equiv\partial_{\mu}\sigma$:

\ba
 R_{\mu \nu}&=&\partial^2\sigma \eta_{\mu
\nu}+2\partial_{\mu}\partial_{\nu}\sigma
-2\sigma_{\alpha}\sigma^{\alpha}\eta_{\mu
\nu}+2\sigma_{\mu}\sigma_{\nu} \nl
R&=&-6e^{2\sigma}(\sigma_{\mu}\sigma^{\mu}-\partial^2\sigma),
\ea
and the lagrange density, which contains new constants ($L_{11},L_{12},L_{13}$),
takes the form:

\ba
&&
[-6(\sigma_{\mu}\sigma^{\mu}-\partial^2\sigma)L_{11}-
(2\sigma_{\mu}\sigma^{\mu}-\partial^2\sigma)L_{12}]<D_{\mu}U(D^{\mu}U)^{\dagger}>\nl
&+&2L_{12}(\sigma_{\mu\nu}+\sigma_{\mu}\sigma_{\nu})<D^{\mu}U(D^{\nu}U)^{\dagger}>\\
&-&6L_{13}e^{-2\sigma}(\sigma_{\mu}\sigma^{\mu}-\partial^2\sigma
)<\chi U^{\dagger}+U\chi^{\dagger}>\nonumber
\ea
The remaining part of the chiral lagrangian is easily reproduced from
(\ref{lag3}) by  exchange of variables: $\gmnd=e^{-2\sigma}\eta_{\mu\nu}$.

Let us proceed to the model evaluations of the structure constants
$L_{11},L_{12},L_{13}$, which come from the low-energy chiral
bosonization method.
The generating functional in the presence of external metric can be reduced
to the following form \ci{Odintsov}:

\be
Z_{\psi}=\int Dq D\bar q\exp(i\int d^4x\,\bar
q\exp(\sigma/2)\wt {\cal D} \exp(\sigma/2)q),
\ee
where the operator   $\wt {\cal D}=\not\!\!{\cal
 D}+e^{-\sigma}(S+iP\gamma_5)\equiv \not\!\!{\cal D}+\wt S+i\wt P\gamma_5$,
with     $ q\equiv e^{-2\sigma}\psi $.
Thus, the conformal metric is induced by local dilatations of the Dirac
operator $q\rightarrow e^{-\sigma/2}q $, and the corresponding effective
action is calculated by integration over the conformal anomaly \cite{Andr2}.

In order to find the chiral lagrangian in external metric we follow the
low-energy QCD-bosonization scheme, presented in the paper \ci{Andr1}.
The collective contribution of the conformal and chiral anomalies into
the Wess-Zumino-type effective action has the following symbolic
form:

\be
W(\Pi,\sigma;\wt S,\wt P)=-\int_{0}^{1}d\tau\,
\mbox{Tr}(\gamma_5\Pi<x|{\cal P}(\Lambda^2-(\Phi_{\tau}\wt
{\cal D}\Phi_{\tau}-iM)^2)|x>) ,
 \ee
where
     $\Phi_{\tau}=\exp((\tau\sigma+i\gamma_5\Pi\tau)/2)$,
and ${\cal P}(...)$
is the finite-mode projector \cite{anbo} 
on the low-energy region in accordance to the model
\ci{Andr3}.

As the result of the chiral bosonization one has the effective
action for pseudoscalar fields, which, in this case, is
expressed by the difference of Wess-Zumino functionals in the external
metric:

\be
S_{eff}(\sigma,\Pi)=W(0,\sigma;\wt S,\wt P)-W(\Pi,\sigma;\wt S,\wt P).
\ee
It can be equivalently expressed by combination of the chiral
lagrangians in  conformal metric (see the section 2) and
conformal ones in the presence of external pseudoscalar
fields, namely:

\ba
&&(W(0,0;\wt S,\wt P)-W(0,\Pi;\wt S,\wt P))\nl
&+&(W(0,\Pi;\wt S,\wt P)-W(\sigma,\Pi;\wt S,\wt P))  \\
&-&(W(0,0;\wt S,\wt P)-W(\sigma,0;\wt S,\wt P)).  \nonumber
\ea
Due to this, one  succeeds to  use the results \ci{Andr2},
where the corresponding conformal effective action was calculated.

In order to convert the model lagrangian
into phenomenological form (\ref{lag3}) ,
we, as before, use the equation of motion:

\be
U^{\dagger}D^2_{\mu}U-(D_{\mu}^2U)^{\dagger}U+4\sigma_{\mu}(D_{\mu}U)^{\dagger}U
+e^{-2\sigma}(\chi^{\dagger}U-U^{\dagger}\chi)=\frac13e^{-2\sigma}
<\chi^{\dagger}U-U^{\dagger}\chi >,
\ee
which, in this case, includes the dependence on the external metric, at that
the tachion-like terms gives the additional contribution  into  the
vertices $L_{11},L_{12},L_{13}$; in particular, the following
vertex becomes essential:

\be
\frac{N_c}{192\pi^2}<(D^2U)^{\dagger}D^2U>.
\ee

After the required transformations we come to the following model
estimates of the structure constants:

\ba
L_{11}&=&1.58\cdot 10^{-3}\nl
L_{12}&=&-3.2\cdot 10^{-3}\\
L_{13}&=&0.3\cdot 10^{-3}.\nonumber
\ea
One should point to the constraint $2L_{11}=-L_{12}$ which has a Zweig
rule form and combines the vertices including curvature tensor in
the conserving  Einstein energy-momentum tensor:

\be
R_{\mu\nu}-\frac12\gmnd R.
\ee

\section{
Conclusion.
}

\hspace*{5mm}In the paper we presented the extended low-energy
bosonization of the vector, axial-vector, scalar and pseudoscalar
$ U(3)$-quark currents, and  also the pseudoscalar gluon density and the
quark energy-momentum tensor.
We have used phenomenological solution of the
$ U(1) $-problem and a model of the low-energy region.
On this way the estimations of the new structure  constants
$L_{11},...,L_{19}$ have been found and Zweig-type constraints
on them were obtained.

It is of interest to compare our model estimations with
    experimental data, available at the moment, and the
 indirect estimates following from the hadron physics.  As it
 was mentioned, the constraint $L_{14}=-4L_7$ is well justified
 by experimental information on the charmonium decays
$\psi'\rightarrow J/\psi$ with emission of the
$ \pi$-or $ \eta$-mesons \ci{Partdat}.

Indeed, in the multipole expansion approach, the branching ratio
of these reactions is given by:

\be
\frac{\Gamma(\psi'\rightarrow
J/\psi\pi^0)}{\Gamma(\psi'\rightarrow  J/\psi\eta)}=
\biggl|\frac{<0|G\widetilde G|\pi^0>}{<0|G\widetilde G|\eta>}\biggr|^2
\frac{p_1^3}{p_2^3}.
\ee
For adopted values of the current quark masses, it makes possible to
find safely the constant $L_{14}$. Experimantal data are precise
enough to evaluate $L_{14}=(2.3\pm1.1)\cdot~10^{-3}$.

For the description of the other structure constants the reliable experimental
information is absent. However, some of them can be evaluated when
proceeding from the phenomenological lagrangian, which includes heavier
hadron states and is reduced to the  chiral lagrangian in the
infrared region \ci{DonLeu}.  In particular, the coefficients
$L_{11},L_{12},L_{13}$ are saturated by the $ \rho $-meson, tensor
and scalar resonances and numerically must be of order:

\ba
L_{11}&=&1.6\cdot 10^{-3}\nl
L_{12}&=&-2.7\cdot 10^{-3}\\
L_{13}&=&0.9\cdot 10^{-3}.\nonumber
\ea
Here one can see a good agreement with evaluations of $L_{11},L_{12}$
and the serious discrepancy of $L_{13}$.
It is entailed by an uncertainty in modelling of the low-energy region
 and serves  as a stimulus for proceeding development
of the chiral bosonization method.

\end{document}